\documentclass 
{aa}

\usepackage{natbib}
\usepackage{psfig}

\newcommand{\apjl}{ApJ Let.}
\newcommand{\apj}{ApJ}
\newcommand{\mnras}{MNRAS}
\newcommand{\nat}{Nature}
\newcommand{\apjs}{ApJ Supp.}
\newcommand{\aap}{A\&A}

\begin{document}

\title{The effect of supernova natal kicks on compact object 
merger rate}

\titlerunning{The effect of supernova natal kicks...}

\author{Krzysztof Belczy{\'n}ski and Tomasz Bulik}
 
\institute{Nicolaus Copernicus Astronomical Center, 
Bartycka 18, 00-716 Warszawa,Poland}

\date{Received , Accepted...}

\thesaurus{02.07.2, 08.02.03, 08.05.03}
 
\maketitle 
 
\begin{abstract} 
{ Mergers of compact objects may lead  to
different astrophysical phenomena: they may provide sources of
observable gravitational radiation, and also may be connected
with gamma-ray bursts. Estimate of the rate with which  such
mergers take place are based on assumptions about various
parameters describing the binary evolution.  The distribution of
one of these  parameters -  the kick velocity a neutron star
receives at birth will strongly  influence the number and
orbital parameters of compact objects binaries. We calculate
these effect using population synthesis of binary stars  and
show that the expected compact object merger rate changes by a 
factor of 30 when the kick velocity varies within its current 
observational bounds. } \end{abstract}


\section{Introduction}

The gravitational wave detectors LIGO and VIRGO will soon be
operational. Their completion brings up the question of the
possible sources of gravitational waves and also their
brightness and observable rate. The most often considered
sources of gravitational waves are mergers of compact objects;
neutron stars or black holes. A number of authors have already
considered these phenomena, and  several estimates have been
published 
\citep{1991ApJ...379L..17N,1991ApJ...380L..17P,1993MNRAS.260..675T,1996A&A...312..670P}.

There are two approaches leading to the calculation of the
merger rate. The first method, let us call it the experimental
approach, is based on the observational fact that we do see
three binary systems of neutron stars: PSR1913+16
\citep{1989ApJ...345..434T}, PSR1534+12
\citep{1991Natur.350..688W}, and PSR2303+46
\citep{1988ApJ...332..770T}. Based on this number, and
considering various  observational selection effects, like e.g.
the pulsar beam width, one can estimate the number of such
systems in our Galaxy. Given the observed orbital parameters for
these systems one can then calculate the lifetime of each one
and thus obtain the expected merger rate in the Galaxy. This
approach suffers from several weaknesses: it is  based on
observations of only three objects, so small number statistics 
strongly affects the results. Moreover, the estimate of the rate
is based on  theoretical assumptions regarding the selection
effects. These assumptions may lead to systematic errors of
uncertain value.

The other approach that we can use, (let us call it a
theoretical one), is based on population synthesis of binary
systems. All compact objects that will finally merge, must have
their origin as ordinary stars in binary systems. 
First studies in that field were performed for Monte Carlo 
simulations of radio pulsars by \citet{1987ApJ...321..780D}.
Thus it seems that by modeling the evolution of stellar 
binaries and the statistical properties of large ensembles of 
binaries one can calculate  the expected population of compact 
object binaries.
Analyzing this population and then considering the systems that
will have  merged within the Hubble time one can also estimate
the present  compact object merger rate in the Galaxy.  This
calculation  requires the supernova rate in the Galaxy,  and
also the fraction of all stars that are in binaries.  This
"theoretical approach" also suffers from the fact that it is
based on several assumptions. In order to create a population of
binaries one requires distributions of initial parameters, like 
the mass of the primary star or the initial mass ratio in the
binary. Some stages in the stellar evolution require
parameterization as well, e.g. mass loss and angular momentum
loss through stellar wind, mass exchange in the common envelope
phase etc. Another parameter that may strongly influence the
number of compact object binaries as well as their lifetime is
the  value of the velocity kick that a neutron star receives as
a  result of supernova explosion.  The importance of this
parameter has been shown by  \citet{1996A&A...312..670P}.

The measurement of the distribution of kicks in supernova
explosions is not easy. One approach that can be taken is to
measure velocities of pulsars and use this distribution as the
distribution of supernova kicks. Here one has to take into
account  selection effects, like for example the fact the  fast
neutron stars may leave the Galaxy and not be visible as  radio
pulsars. A comprehensive study of this and other selection
effects \citep{1997AAS...19111308A} shows that  a large
fraction  of neutron stars has velocities above $500$\
km~s$^{-1}$. \citet{1996ApJ...456..738I} argue that the
velocities of radio pulsars can be explained by just taking into
account the recoil velocity due to mass loss in supernova
explosions in binaries, and neglecting a possible "natal kick"!
An independent study of velocities of young pulsar through
measuring offsets from the centers of supernova remnants 
\citep{1994ApJ...437..781F,1994Natur.369..127L} confirms  that
the distribution initial velocities of neutron stars may have a
large high velocity tail.   On the other hand
\citet{BlauwRama1998} argue that the observed properties of
pulsars can be explained in a model with a single unique value of
the kick $v_{kick}= 200$\,km~s$^{-1}$.  A similar result has been
obtained by \citet{1997MNRAS.288..245L} who show that the
observed population of neutron stars is consistent with the kick
velocities in the range $150\, - \, 200$~km~s$^{-1}$. The
distribution of the kick velocities has been parameterized by
different authors: \citet{1996A&A...312..670P} used a Gaussian
with the width of $450$\ km~s$^{-1}$, while \citet{1997ApJ...482..971C}
proposed a weighted sum two Gaussians: 80 percent with the width
$175$\ km~s$^{-1}$ and 20 percent with the width
$700$~km~s$^{-1}$. Here we attempt to investigate systematically
the effect of the  kick velocity distribution on the compact
objects merger rate. In section~2 we describe the model of
binary evolution used for simulating the population of binaries,
in section~3 we show the results of the simulation  and
calculate the compact object merger rate. Finally we discuss our
results in section~4.

\section{Population and evolution of binary systems}

While the evolution of a single star is only a function of its
mass and metallicity the evolution of a binary presents a more 
complicated problem. It depends on the mass of a more massive 
component, the mass ratio of the smaller to the larger star
$q$,  and on the initial parameters of their orbit: $e$ - the 
eccentricity and $a$ the semi-major axis of the orbit. 
Following the general approach in the field we assume that these
parameters are independent, i.e. the probability density
can be expressed as a product: $p(M,q,e,a) =
\Psi(M)\Phi(q)\Xi(e)\Gamma(a)$.
 Our
population synthesis code is mainly based on  \citet{Bethe1998}.
The evolution of a single star  is described by
\citet{1989ApJ...347..998E}. In the following all masses are in
the units of solar mass $M_\odot$.

\subsection{Distribution of the initial parameters}

Beyond $10M_{\odot}$ the distribution of the masses of the
primary  stars that we use follows \citet{Bethe1998},
\begin{equation}
\Psi(M) \propto {M}^{-1.5}
\end{equation}
Other authors have used similar distributions, e.g. 
\cite{1996A&A...309..179P} use the exponent of $-1.7$. 
We restrict the range of the masses to the interval 
between $10M_\odot$ to $50M_\odot$ as we want at least 
the more massive star to undergo a supernova explosion.

We adopt the following distribution of the mass ratio $q$
\begin{equation}
\Phi(q) \propto {\mathrm {const}}
\end{equation}
There is some uncertainty regarding $\Phi$.
\citet{1996A&A...309..179P} use $\Phi\propto (1+q)^{-2}$, while 
\citet{1994MNRAS.268..871T} use $\Phi\propto q$.  
Yet another distribution is used in the simulation
by \citet{1994A&A...288..475P}, who use a distribution 
slightly peaked for small values of q and falling down 
when $q$ goes to unity.

Initial binary eccentricity $e$ is assumed to have value between 0 
and 1 and its initial distribution is taken from 
\citep{1991A&A...248..485D} :
\begin{equation}
\Xi(e) = 2e.
\end{equation}

The distribution of the initial semi-major axis
$a$ used in population synthesis codes is 
flat in the logarithm, i.e.
\begin{equation}
\Gamma(a) \propto {1\over a}.
\end{equation}
Both stars are initially massive main sequence stars, with  the
radii of at least 4--5 R$_{\odot}$,  so we take minimum value of
$a$ to be 10 R$_{\odot}$. However if the sum of radii of two
zero age main sequence stars  exceeds  10 R$_{\odot}$, we
chose the minimal  separation to be twice the radius of the
primary component.   This should give the stars the space to
evolve without merging  before they go through their main
sequence lifetime. Nevertheless some systems might be  born with
high eccentricity, and thus be in contact and merge forming a
single very massive star. At this point  we stop our
calculation  as we are not interested in this type of mergers in
our study.

Although very wide binaries with periods up to 10$^{10}$\ days
are  observed \citep{1991A&A...248..485D} and models with
initial separation  up to 10$^{6}$ R$_{\odot}$ are used
\citep{1996A&A...309..179P}, we  have set the maximal separation
to 10$^{5}$ R$_{\odot}$ as wider binaries  are very unlikely to
survive two supernova explosions and form  a compact object
system.  Our models showed that setting the upper limit to
10$^{6}$ R$_{\odot}$  does not change the  results noticeably.

We assume that the distribution of the kick velocity a newly 
born neutron star receives in a supernova explosion is a three 
dimensional Gaussian and parameterize it by its width
$\sigma_v$.  We  vary $\sigma_v$ to determine  the dependence of
the merger rate and properties of the compact object binaries
population on this parameter. Alternatively, we draw the natal
kicks from a weighted sum two  Gaussians
\citep{1987ApJ...321..780D}: 80 percent with the width $175$\
km~s$^{-1}$ and 20  percent with the width $700$~km~s$^{-1}$.

\subsection{Evolution of binaries}

In our calculations we neglect the effects of stellar wind on
evolution of binaries.

Tidal circularization takes place in binaries for which the size
of any component (or both components) is large in relation  to
their separation\citep{1978A&A....67..162Z}. This  happens when
the stars evolve to the giant stage,  and/or if the initial
binary separation is small.  We follow the prescription proposed
by \citet{1996A&A...309..179P};  the circularization takes place
if the stellar radius of one component  is larger then 0.2 of
the  periastron binary separation.  The orbital elements ($a,e$)
change  with conservation of  angular momentum until the new
binary separation is 5 stellar radii  of the component in which
tidal effects take place, or the binary is  totally circularized
($e=0$).

Binary stars may undergo 
mass transfer and common envelope phases in different 
evolutionary stages. Depending on the physical conditions this
will result in the change of mass of each star, mass loss from
the system,  and also  in change of the size and shape of the
orbit.  We describe below three schematic evolutionary paths
and  accompanying mass transfer types used in the code.

\subsubsection{ Evolution for  small mass ratios,
$(q<0.8897)$}

The more massive star (primary) evolves faster in the binary. 
We use the following approximate formula for the  main sequence
lifetime of  a star with mass $M$: $T_{\rm MS} = 20 \times
10^{6}\ (M/10M_{\odot})^{-2}$\ yrs.  After this time the primary
evolves to the giant stage which   lasts about 20 per cent of
its main sequence lifetime, and  increases its size. For
$q<0.8897$ the secondary   is still on main sequence at the end
of primary giant stage.  If the giant primary overfills its
Roche lobe mass transfer to the  main sequence companion (star
2, with mass $M_2^i$) and mass  loss takes place. We denote this
regime by {\em type I} mass transfer.  The mass transfer continues until the
giant is stripped of its  envelope, and  thus the final mass of
the giant will be just that of the its  helium core. We use the
approximation $M_1^f = 0.3\, M_1^i$ \citep{Bethe1998}. A part of
the envelope mass $0.7\, M_1^i$ is lost from the system   while
another part is transferred to the companion.  The fraction of
mass transferred to the companion is proportional  to the square
of the mass ratio   \citep{1991A&A...249..411V,Bethe1998}, so
the mass of the companion  after the mass transfer is 
\begin{equation}
M_2^f =
M_1^i (q + 0.7 q^2)\, ,
\label{mfin1}
\end{equation}
where $q$ is the initial mass ratio. 

The orbit was already circularized by tidal forces, when the  
giant was filling its Roche lobe, and now its size changes
due to mass transfer and mass loss.
We describe the change in semi-major axis by
\citep{1994A&A...288..475P} 
\begin{equation}
{a_f \over a_i } = \left( {M_1^f \over M_1^i} {M_2^f \over M_2^i} \right)^{-2} 
                \left( {M_1^f + M_2^f \over M_1^i + M_2^i} \right)^{2\beta+1}, 
\end{equation}
where $a$
 is binary separation, $M_1$ is the mass of the giant 
losing material, $M_2$ is the mass of its main sequence companion 
and indices $i,f$ correspond respectively  to the values 
before and after the Roche lobe overflow.  
The parameter $\beta$ describes the specific angular momentum of material 
leaving binary \citep{1994A&A...288..475P}. 
The above equation assumes that $\beta$ is constant with time.
The value of $\beta$ is uncertain; typically 
the values  such as $\beta \geq 6$ or  $\beta = 3$ are used, for 
 discussion see  \citet{1994A&A...288..475P}.
Here all calculations are performed for $\beta=6$.    
The initial primary, now either a helium star (after the 
mass transfer) or a giant goes supernova, leaving behind a 
neutron star.
The initial secondary evolves, becomes a giant and may begin to 
transfer material to the newly formed neutron star increasing its 
mass, and possibly  turning it to a black hole.
We assume that the maximum mass of a neutron star is  
$2.2 M_{\odot}$.

In describing accretion on a neutron star we follow
\citet{Bethe1998}, and we will call this
regime {\em type II} mass transfer. 
A neutron star accretes matter while moving in  orbit through the 
extended envelope of the giant. 
We use the  Bondi-Hoyle accretion rate 
\begin{equation}
\dot M = \pi \rho v R_{ac}^2
\end{equation}
where $\rho$ is the density in the vicinity of the compact
object of mass $M_1$, $v$ is its velocity, and 
$R_{ac} = 2G M_1 / v^2$ is the accretion radius. 
Consideration of the energy loss equations lead to the final mass of 
the compact object \citep{Bethe1998}
\begin{equation}
M_1^f = 2.4 \times\left( M_2^i \over M_1^i \right)^{1/6} M_1^i\,
.
\label{m2f1}
\end{equation}
We assume that the giant loses its envelope so its final mass is
just that of the helium core $M_2^f = 0.3\, M_2^i$. 
The orbital separation  follows from the energy integration i.e.
equations 5.18 through 5.23 in
\citep{Bethe1998}:
\begin{equation}
a_f = 0.145\times a_i \left( {M_1^i\over M_2^i} \right) \, .
\end{equation}
Finally the initial secondary undergoes a supernova explosion,
and provided that 
the system survives, we obtain a compact object binary, 
consisting of either two neutron stars or 
of a black hole and a neutron star.

\subsubsection{ Evolution of two stars for intermediate 
mass ratio, $(0.8897<q<0.95)$.}

When $q>0.8897$ the secondary is already
a giant when the primary explodes as a supernova.
The upper limit is explained in the next subsection.

For the  case of small orbital separations the evolution  begins
in a similar way as for {\it small} $q$ described above.  The
evolution begins in a similar way as in the case of small mass
ratio, and the primary may transfer mass to the secondary as
described by {\em type I} above. When the secondary becomes a
giant the primary is already a helium star, stripped of its
envelope. The secondary may transfer mass to the primary 
and we approximate this process also by {\em type I}.
The primary explodes as a supernova and forms a neutron star,
which may accrete from the secondary, provided that it is 
still a giant. The mass transfer in this phase is 
described by {\em type II}. The secondary undergoes a supernova
explosion and if the system survives we obtain a compact object 
binary.

In the case of  large orbital separations the first stage of 
mass transfer from the giant to the main sequence star (type I) 
does not occur.  The stars might begin to interact only when
both of them are  already in the giant stage.  In this case they
loose the common envelope. To describe this stage of evolution 
we assume that the giants loose the 
entire envelopes and become helium stars of 30 per cent of the
initial  giant masses.
We then approximate the change in the size of the orbit by: 
\begin{equation}
{a_f \over a_i } \approx \left[ 1 + {(M_1^i + M_2^i) M_1^i (1 -
q^2)\over 
        \alpha_{CE} M_1^f M_2^f } \right]^{-1}.
        \label{ce}
\end{equation} 
As above $a$\ is binary separation, $M_1, M_2$\ are the  masses
of  primary and secondary. 
The indices $i,f$ correspond  to the values before and 
after common envelope phase, respectively, and $\alpha_{CE}$
describes the efficiency of the orbital energy expenditure 
in the dispersal of common envelope.
Equation~(\ref{ce}) was adopted from \citet{1993YT} 
and describes the change of the orbital separation 
$a$ following decrease of the orbital
energy which was used to eject a common envelope.
They have also showed that $0.6 \leq \alpha_{CE} \leq 1$, and we 
have performed our calculations for different values: 
$\alpha_{CE}=0.4, 0.6, 0.8, 1.0$. 
Our calculations show that the results (the production rate
of the compact object binaries, and  their properties) are almost
indistinguishable for these three values, so for simplicity we 
present our results for one value of $\alpha_{CE}=0.8$.

Similarly to the case of small orbital separations
if the secondary is still a giant it may transfer mass to
the neutron star formed in the supernova explosion of the
primary. This mass transfer is described as {\em type II}.

\begin{figure}

\psfig{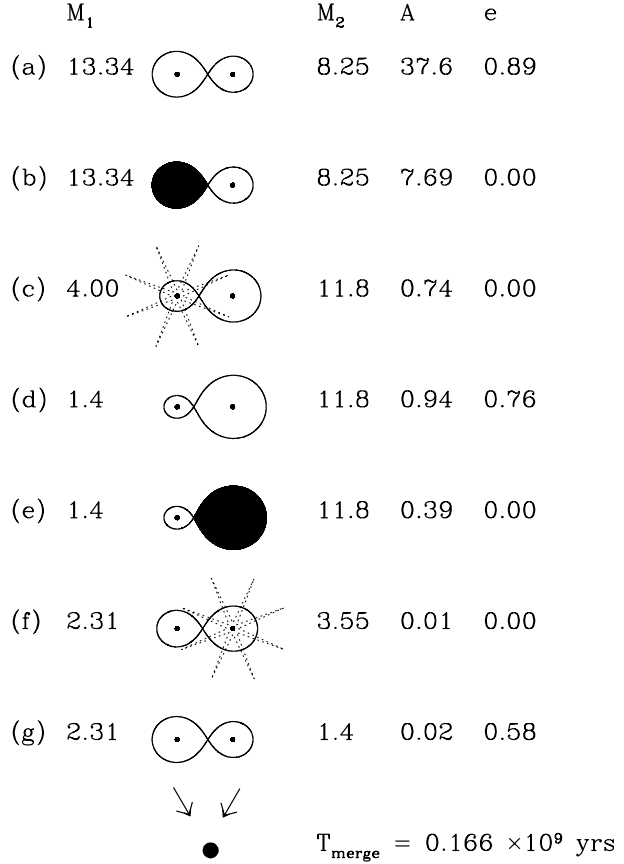}

\caption{An example evolutionary path leading to
formation of  a black hole neutron star binary. Units of mass are
solar masses, and units of distance are astronomical units:
(a) initial binary; (b) after tidal circularization when star 1 expanded to
giant size; (c) system after first mass transfer, star 1 explodes as
supernova; (d) system after first supernova; (e) after tidal circularization
when star 2 expanded to giant size; (f) system after second mass transfer,
star 2 explodes as supernova; (g) system after second supernova: black hole
neutron star binary.
}
\label{evol1}
\end{figure}

\begin{figure}
\psfig{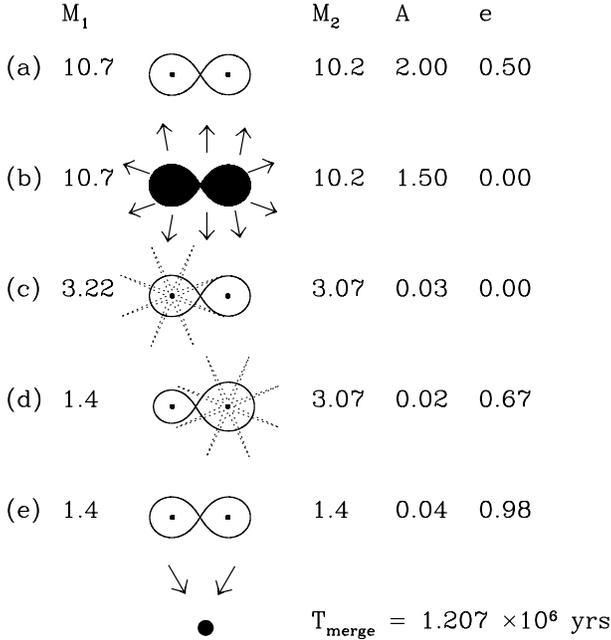}

\caption{An example evolutionary path leading to
formation of a neutron star neutron star
binary
formation:
(a) initial binary; (b) after tidal circularization when both
stars expanded
to giant sizes, system ejects common envelope; (c) system after
common envelope
phase, star 1 explodes as supernova;
(d) system after first supernova, now star 2 explodes as
supernova;
(e) system after second supernova: neutron star neutron star
binary.
}
\label{evol2}
\end{figure}

\begin{figure*}
\psfig{width=0.95\textwidth,file=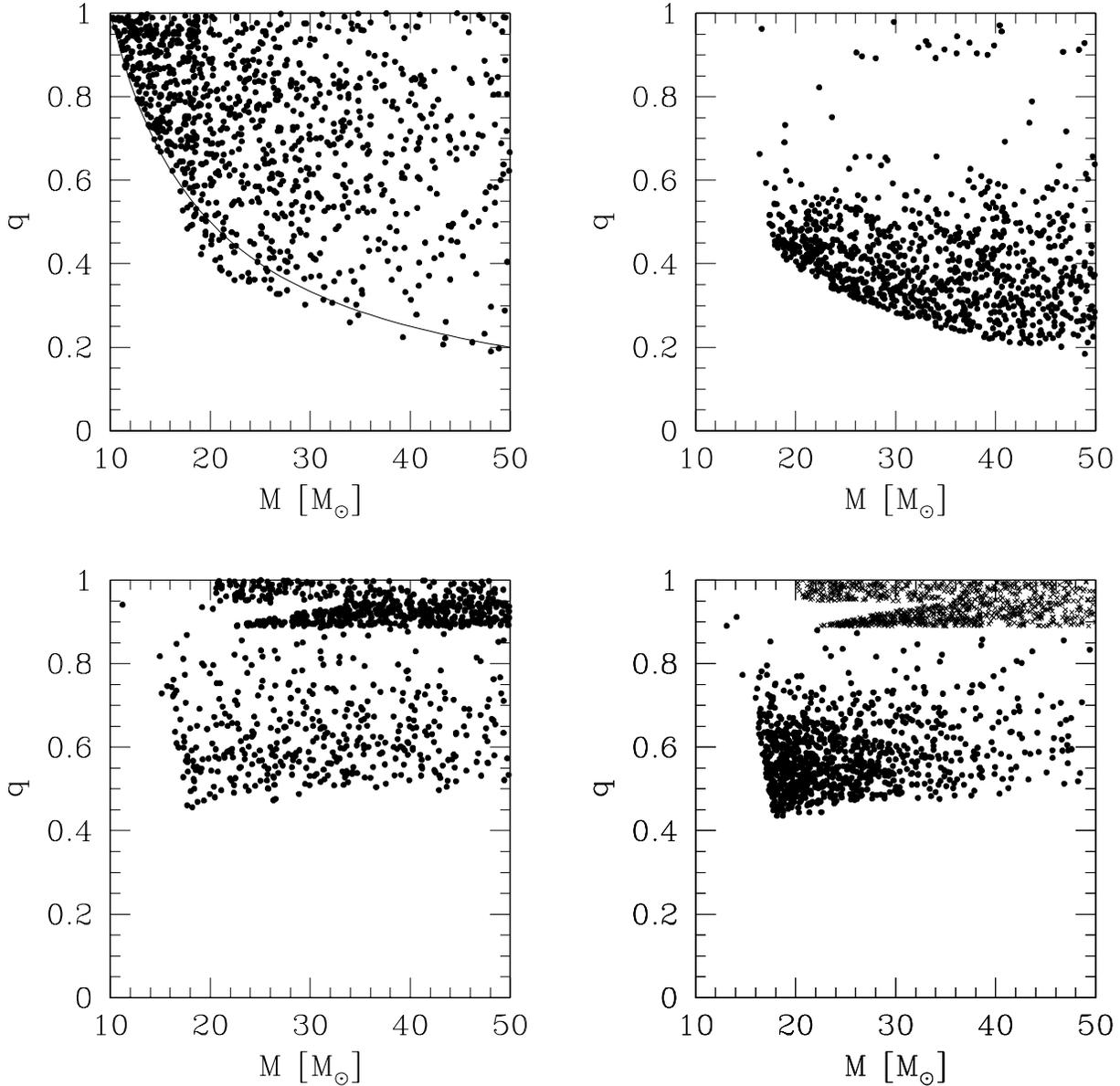}
\caption{Possible results of the evolution of a binary 
with the initial orbital separation $a=20$R$_\odot$ and
eccentricity $e=0.5$ for different values of the initial primary mass $M$
and the initial mass ratio $q$. 
The top left panel shows the systems that 
were disrupted in the first supernova explosion.
The solid line corresponds to the initial secondary mass 
$10\,M_\odot$. The top right panel shows the systems
in which the neutron star, born in the first supernova
explosion, merged with the companion.
The bottom left panel shows the systems disrupted in the second
supernova explosion. The crosses in the bottom right panel
show the neutron star binaries and the filled circles are the 
black hole neutron star binaries. We present one thousand of each
type of binaries. In this calculation we used the Cordes \&
Chernoff (1997)
kick velocity distribution.
}
\label{e5a20}
\end{figure*}

\begin{figure*}
\psfig{width=0.95\textwidth,file=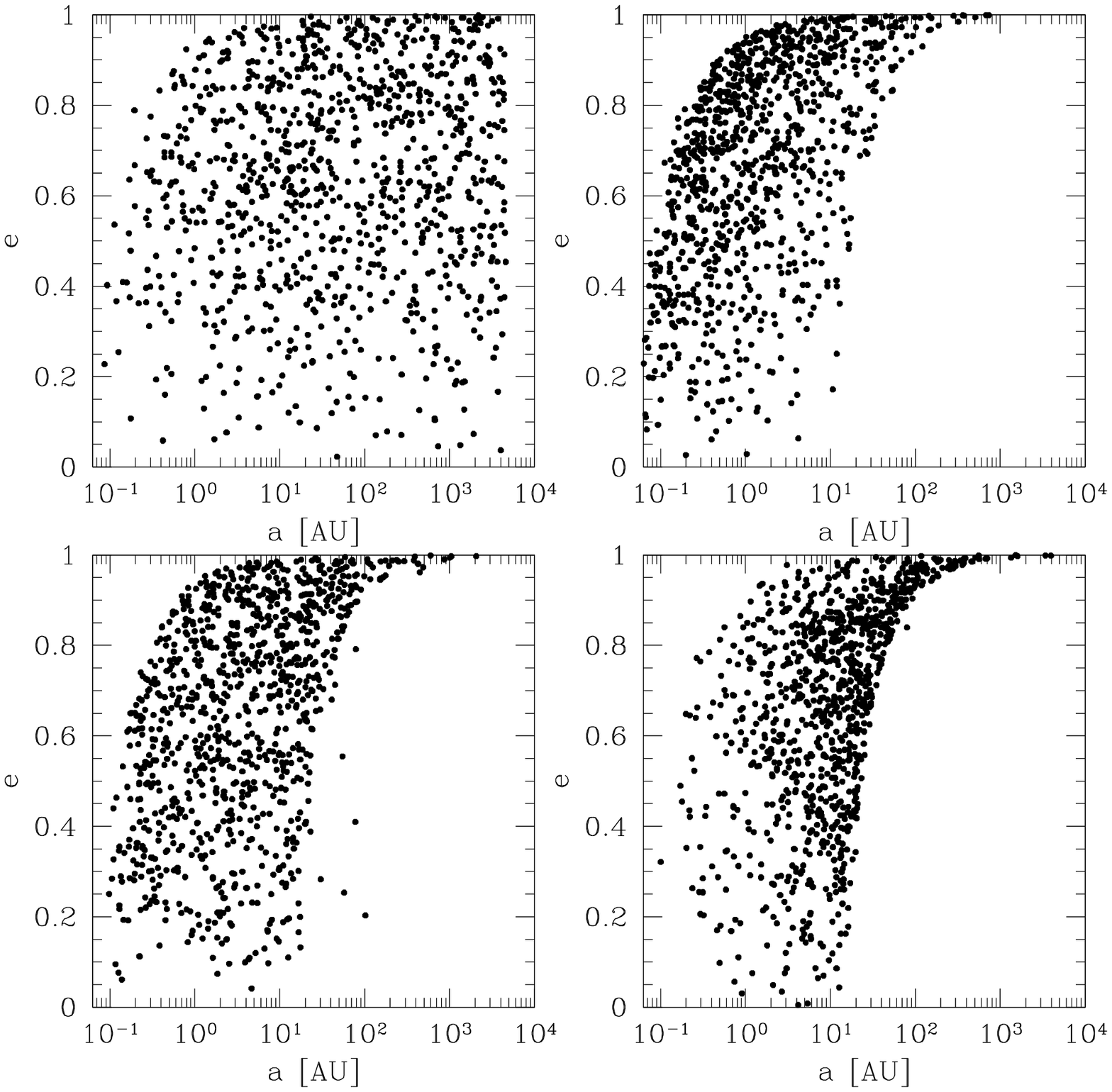}
\caption{Possible results of the evolution of a binary 
with the initial primary mass $20\,M_\odot$ and the mass ratio
$q=0.5$. The top left panel shows the systems disrupted in the
first supernova explosion, the top right panel shows the systems
in which the neutron star, born in the first supernova
explosion, merged with the companion.
The bottom left panel shows the systems disrupted in the second
supernova explosion.  The bottom right panel shows
the black hole neutron star binaries, and there are no
neutron star binaries formed for this mass ratio ($q=0.5$).
We present one thousand of each
type of binaries. 
In this calculation we used the Cordes \& Chernoff
(1997)
kick velocity distribution.}
\label{m20q5}
\end{figure*}

\subsubsection{Evolution of two stars of almost equal mass,
$(q>0.95)$}

Two stars almost simultaneously (within $10$\% of their main 
sequence lifetime) leave main sequence and become giants.  If
their radii are large enough in relation to the periastron 
orbital separation, the orbit is tidally circularized. If the
binary separation is small enough  the giants  may finally
overfill their Roche lobe and enter a common envelope  phase,
which we describe by equation~(\ref{ce}). The two stars go
supernova one after another, and if the system  survives, we
obtain a compact object binary consisting of  two neutron stars.

After each  process which changes the orbital parameters and
sizes of the components  we check if the components sizes are
not larger than the periastron  binary separation.

\subsection{Supernova explosion}

The supernova explosions are very likely to have an impact on the
binary systems. We assume that in each supernova explosion a
neutron star with mass of $1.4\, M_\odot$ is formed and we neglect 
the interaction of the companion star with the envelope ejected in 
the supernova explosion.

We draw a random time for a supernova explosion in the orbital 
motion of the binary. 
We find the relative position and velocity of the two stars 
for this moment.
The exploding star is then replaced by a neutron star and we 
add the kick velocity to its orbital velocity, while the
expanding supernova shell carries away a part of the total
momentum of the binary.
If the energy of such a system is less than zero the system is 
bound and we calculate the parameters of a new orbit of the 
binary and its new spatial velocity.  
For a detailed discussion of the effect of sudden mass loss and a 
random kick velocity on binary parameters see  \citet{1983ApJ...267..322H}.  

After the first supernova explosion we verify if
the neutron star does not collide and merge with 
the companion, i.e. if the radius of the companion is larger
than the periastron binary separation.

\subsection{Energy loss through gravitational radiation}

Once a compact object binary is formed its orbit will evolve 
because of the gravitational wave energy loss.
Orbital energy loss through radiation of gravitational waves 
becomes important once a compact object binary is in a tight 
and/or
highly eccentric orbit. The evolution of  the
semi-major axis $a$ and eccentricity of the orbit $e$, in a
binary emitting gravitational waves have been calculated by
\citet{Peters1964}. The lifetime of a compact object binary is 
\begin{equation}
t_{merg}= {5c^5a^4 (1-e^2)^{7/2} \over 
256G^3 M_{1}M_{2} (M_{1}+M_{2})} \left( 1+ {73\over 24} e^2 + {37\over 96}
e^4\right)^{-1}
\end{equation}
where $a$ is the semi-major axis of the orbit,
 $e$ is its eccentricity, and $M_{1},\, M_{2}$ are the masses of the
 compact objects.

\begin{figure}
\psfig{width=8cm,file=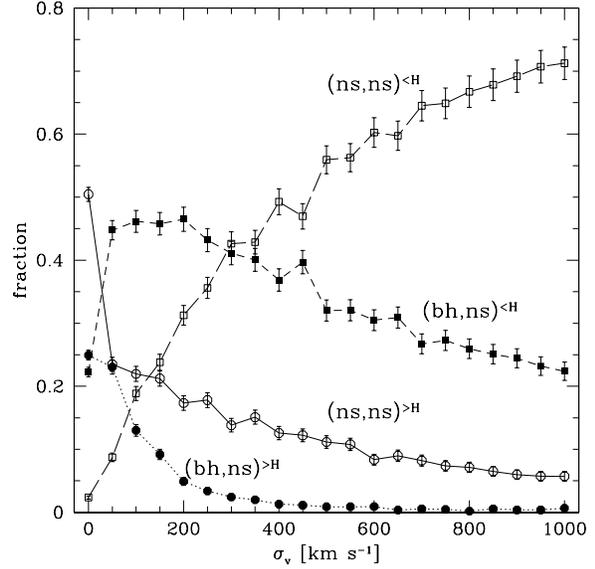}
\caption{Population of the compact object binaries 
in the four categories: double neutron star  binaries that merge
within the Hubble time (empty squares), and these that do not 
(empty circles), neutron star black hole binaries that merge 
within the Hubble time (filled squares), and these that do not 
(filled circles). The error bars represent the counting
statistics of the simulation.
}
\label{frac}
\end{figure}

\section{Results}

\begin{figure*}
\psfig{width=0.95\textwidth,file=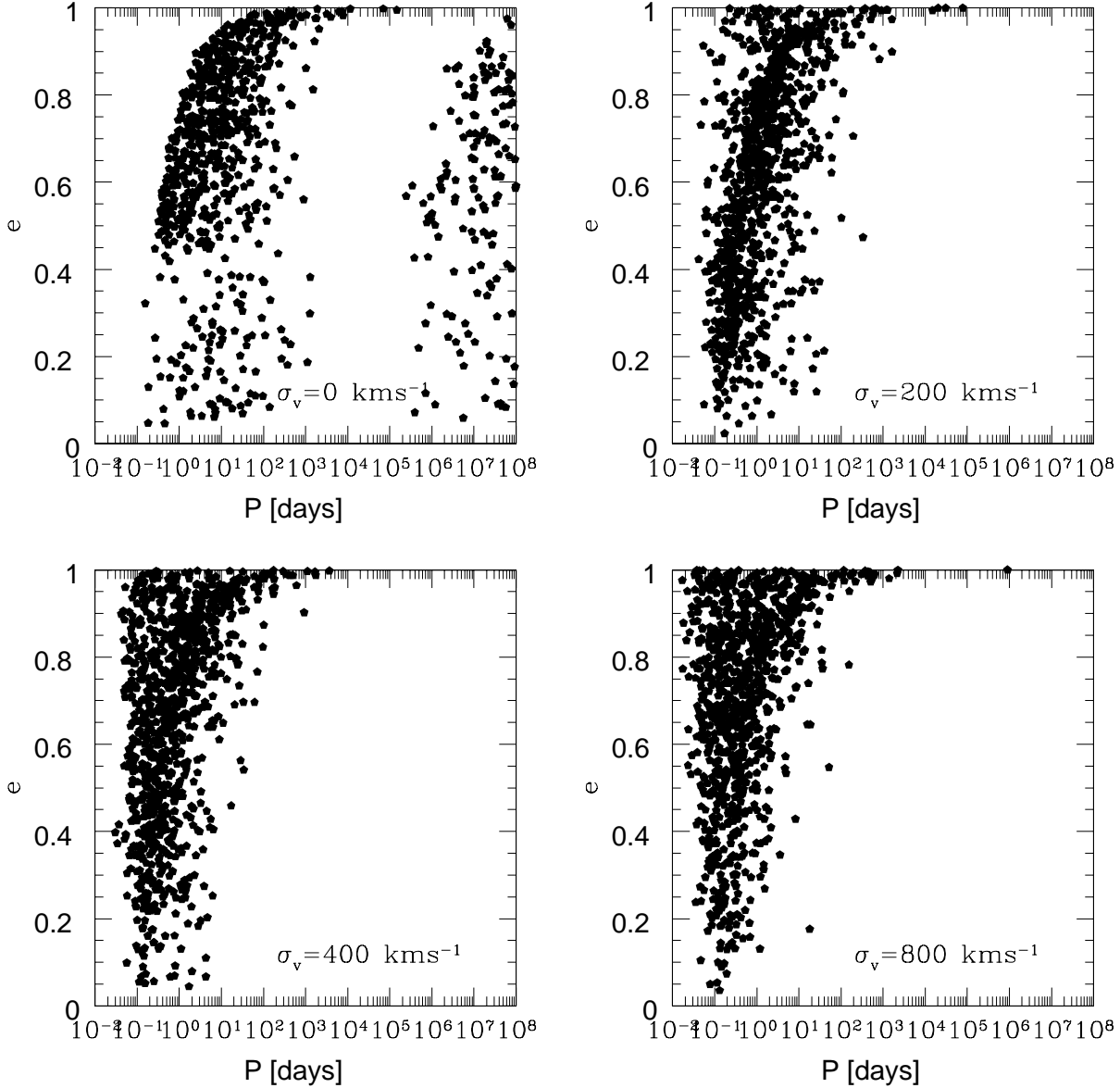}
\caption{Distribution of the initial orbital compact binary 
systems parameters in the space spanned by the period  $P$ 
and eccentricity $e$, for four different Gaussian distributions 
of the kick velocity 
$\sigma_v=0$~km~s$^{-1}$ - top left panel,  $200$~km~s$^{-1}$ -
top right panel, $400$~km~s$^{-1}$ - bottom left panel, and
$800$~km~s$^{-1}$ - bottom right panel. Each panel contains 1000
objects.}
\label{pefour}
\end{figure*}

The code  described above allows to generate populations of
compact object binaries and trace their statistical properties.
In this paper we mainly concentrate on the dependence of these
properties as a function of the  kick velocity distribution.  In
Figures~\ref{evol1} and~\ref{evol2} we present two example
evolutionary paths  leading to formation of compact object
binaries: black hole neutron  star binary and
double neutron star systems.

Let us first consider different evolutionary paths of the
binaries depending on their initial parameters. We first consider
a model with the \citet{1997ApJ...482..971C} kick velocity
distribution. Tracing the evolution now depends on four 
parameters: $M$ (primary initial mass), $q$ (initial mass ratio), 
$a$ (initial semi-major axis), and $e$ (initial eccentricity). 
In order to visualize the
evolutionary effects we fix two of them and present the types of
systems obtained in the course of the binary evolution. In
Figure~\ref{e5a20} we fix the initial orbital separation
$a=20\,R_\odot$ and eccentricity $e=0.5$. The graphs are empty in
the lower left part for which $M\times q< 10M_\odot$. This is
the region for which the secondary is not massive enough to
undergo a supernova explosion. However, in some systems  for
which the primary mass is $M>20M_\odot$, the mass of the
secondary is increased by accretion when  the primary goes
through the giant phase. Most of the systems end up either by
disruption in the first supernova explosion or by a spiral in of
the neutron star to the secondary (top panels) The remaining
systems may be disrupted in the second supernova explosion. The
compact object binary population (bottom right panel) is
bimodal. The neutron star  neutron star binaries are formed from
the systems with $q$ very close to unity, while  the black hole
neutron star systems are primarily formed  when $q$ is
intermediate.  Thus the initial distribution of the mass ratio
$q$ has a strong influence on the production rates of these two
types of compact object binaries.  The number of systems shown in
each panel does not correspond to the actual production rates.
We present one thousand systems in each panel.
The actual calculation produces 51\% systems with the secondary not
massive enough to undergo a supernova explosion, 
40\% systems torn after the first explosion, 
6.2\% of systems merged after the first explosion, 
2.7\% systems torn after the second explosion, and 0.35\% of 
compact object binaries.

In Figure~\ref{m20q5} we present the results of the evolution of
systems for which we fix the initial primary mass and the
initial  mass ratio, while varying the the initial orbital
parameters $a$ and $e$. The systems with small orbital
separations and high eccentricity merge in the early phase of
the evolution and are not considered in this paper.  Disruption
after the first supernova explosion may occur to all  systems. 
The population of systems that end up merging after the first
explosion, are disrupted in the second explosion, or form a
compact object binary originate from the same region of the
parameter space. One should note that because of the
circularization of orbits already in the initial stages of the
binary evolution the final population is not very sensitive to
the initial eccentricity. On the other hand the distribution of
the initial orbital separation is important, as the compact
object binaries originate in systems with  relatively small
orbital separations (see bottom right panel in
Figure~\ref{m20q5}). One should note that all the compact object
binaries shown in Figure~\ref{m20q5} are black hole neutron star
binaries. Double neutron star systems are formed only when $q$
is nearly unity in our simulations. As before the number of
systems shown in each panel does not correspond to the actual
production rates, and show one thousand systems in each  panel. 
The actual calculation produces 12\% systems born in contact,
43\% systems with the secondary not massive enough to undergo a
supernova explosion, 34\% systems torn after the 1st explosion,
0.7\% systems merged after the  first explosion, 0.50\% systems
torn  in the second explosion and only 0.15\% mergers.

We present the dependence  of the production rates of different
types of compact object binaries on the width of the kick 
velocity distribution $\sigma_v$ in Figure~\ref{frac}. The
number of double neutron star systems that merge
within the Hubble time 
increases with the kick
velocity, while the production rate
of black hole neutron star systems becomes smaller. 
These two rates are nearly equal when the kick velocity is
roughly that given by \citet{1997ApJ...482..971C}.

In Figure~\ref{pefour} we present the distributions of the
orbital parameters of compact object binaries for a few
representative values of the width of the kick  velocity
distribution $\sigma_v$. Each panel contains one thousand
compact object binaries.  To understand these plots let us first
consider the properties of the population of objects just before
the second supernova explosion in the case
$\sigma_v=0$\,km\,s$^{-1}$. Some of the systems are wide, with
the eccentricity varying from zero to unity, however, most of
the systems populate a region  with eccentricities above
$\approx 0.45$. 
These systems originate in binaries with small initial mass 
ratios.
A characteristic property of this group is a
correlation between eccentricity  and period.  Objects in this
group originate in systems with small initial value of $q$.  The
masses of the primary  compact objects in  these systems have a
narrow distribution, the mass of the secondary  after the
accretion phase weakly depends on the initial mass before
accretion - see equation~\ref{m2f1}, and is typically $
{M'}_{2}\approx 2.3\, M_\odot$.  The mass of the secondary just
before the second supernova explosion,  is the mass of the
helium core of secondary star and  is in the range $ 3.0\,
M_\odot<{M'}_{2} <5.5\, M_\odot$. 
The orbital parameters of such system  after an explosive mass
loss (second supernova explosion) depend only on the total mass
loss, see e.g. \citet{1994A&A...288..475P}.  In our calculation
the newly formed  neutron star has a mass of $1.4\, M_\odot$,
thus the relative mass loss $x =(1.4+{M'}_{2})/({M'}_{2}+{M'}_{2})$  
is in the range $0.47<x<0.69$. Systems that loose more than
half of the mass ($x<0.5$) are unbound. Other systems become
eccentric and the eccentricity is given by $e' = (1-x)/x$. Hence
the lowest eccentricity  a system can have after the second
supernova explosion in our simulations is $e' = 0.44$.  The
relation between the new orbital period  $P'$ and the new
eccentricity $e'$ for different relative mass loss $x$ is 
\begin{equation} 
P = P_0 {(1+e')^{1/2}\over  (1-e')^{3/2}}\, , 
\end{equation} 
where $P_0$ is the orbital period before the
explosion. This relation defines the curved shape of the
distribution  in the $P,e$ diagram. The objects populating the
right hand side of the $P,e$ plot, originate from systems with
intermediate  or large $q$.  The intermediate $q$ objects went
through accretion  onto the neutron star  (regime II of a mass 
transfer described above) and therefore similar  reasoning as above 
applies to them. 
However systems with high initial value of $q$  results in wide 
binaries (periods larger than 100 days) with  different 
eccentricities.
The compact object binaries in Figure~\ref{pefour} (top left panel) 
with eccentricities below 0.45 and orbital periods smaller then 
$10^3$\ days originate in binaries of intermediate and large initial 
mass ratio.

In the case of non zero kick velocity the systems  with high $q$
are very unlikely to survive. In fact there are only two such
systems on the plot for $\sigma_v = 200$km~s$^{-1}$, and none
for higher velocities.  The escape velocity in long period
systems  is low, and if such  systems existed after the 
first supernova event, they  have are very likely disrupted in
the second supernova explosion.

\begin{figure}
\psfig{width=8cm,file=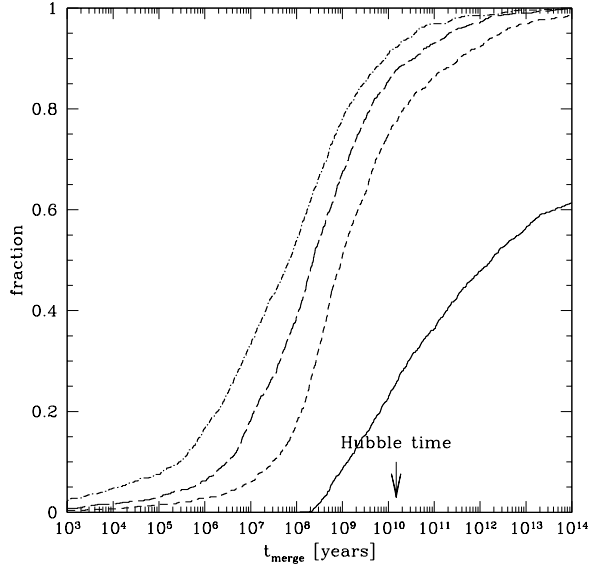}
\caption{Cumulative distributions of the $t_{mrg}$ - the
lifetimes of compact object binaries for four different
kick velocities: the solid line for $\sigma_v = 0$~km~s$^{-1}$,
the short dashed line for 
$\sigma_v = 200$~km~s$^{-1}$,
the long dashed line for $\sigma_v = 400$~km~s$^{-1}$,
and the dash dotted line for $\sigma_v = 800$~km~s$^{-1}$.
Each line was plotted from a distrbution of 1000 objects.
}
\label{temerge}
\end{figure}

As the kick velocity is increased the shape of the distribution
in the  $P, e$ diagram also changes. In the case $\sigma_v =
200$~km~s$^{-1}$ there is a small fraction of high eccentricity,
low period systems. This region of the parameter space fills up
as the kick  velocity increases, see lower panel in
Figure~\ref{pefour} where the case $\sigma_v=400$~km~s$^{-1}$
and $\sigma_v=800$~km~s$^{-1}$ are shown. This is due to the
fact that with the increasing kick velocity the long period
systems are easier to disrupt and the fraction of surviving
short period systems becomes larger.  We note that  the
distributions shown in Figure~\ref{pefour} are similar to those
obtained previously
\citep{1996A&A...312..670P,1993MNRAS.260..675T}.

In Figure~\ref{temerge} we present the cumulative  distributions
of the lifetimes of compact object binaries for the same set of
kick velocities as in Figure~\ref{pefour}. In the case of no
kick velocity $\sigma_v =0$~km~s$^{-1}$  the distribution is
bimodal: the systems with small $q$  merge typically within the
Hubble time (which we take to be $15$\,Myr). However the systems
with higher $q$ remain in wide orbits and their merger times
exceed the Hubble time. With the increasing kick velocity only
the tightly bound systems survive. The lifetime of a system
scales with the fourth power of the semi-major axis, and
therefore the median  lifetime decreases with increasing kick
velocity. In the case of $\sigma_v = 200$~km~s$^{-1}$  the
median lifetime is $\approx 4\times 10^8$~years, and it
decreases by a factor of four when the kick velocity is doubled.
For the highest kick velocity  velocity  $\sigma_v =
800$~km~s$^{-1}$ the median lifetime  is only  $\approx 3
\times 10^7$~years. One should note however, that the
distribution of $t_{\rm merge}$  is very skewed, and we present
it on a logarithmic axis.  Thus there is a long tail extending
to about the Hubble time. Most of the mergers take place in the
Hubble time and only a small fraction of the total population
lasts longer.

\begin{figure}[t]
\psfig{width=8cm,file=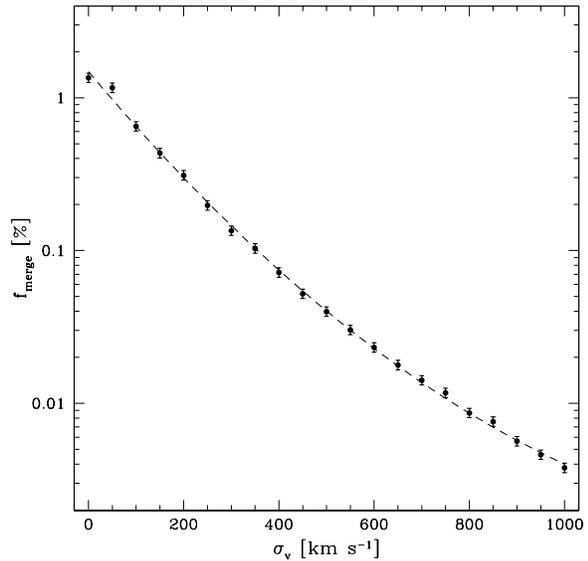}
\caption{Fraction of binary compact objects that merge within 
the Hubble time (15Gyrs), as a function of the width of the kick
velocity distribution. We also show the
fit discussed in the text. Each calculation produced 1000
merging binaries.}
\label{fmerge}
\end{figure}

Finally in Figure~\ref{fmerge} we present $f_{merge}$ -- the
fraction of all binaries with the primary star more massive than
$10M_\odot$ that produce a pair of compact objects in a binary
system. We calculate $f_{merge}$ for a large range of the kick
velocity distribution widths.  To calculate each point in
Figure~\ref{fmerge} we calculated one thousand of compact
binaries. This fraction falls down very approximately
exponentially with the increasing  kick velocity. We have fitted
a modified exponential  to this dependence 
\begin{equation}
f_{merge} \propto
\exp(-4.21 - 8.51\times 10^{-3}\sigma_v +2.6\times 10^{-6}
\sigma_v^2) \, .
\label{fit}
\end{equation}
The fit is shown by the dashed line in
Figure~\ref{fmerge}  and is accurate to about 6\%. 
Note that equation~(\ref{fit}) can be used
to determine the merger fraction for any kick velocity
distribution that can be expressed as a linear combination of
three dimensional Gaussians. Equation~(\ref{fit}) can be used
together with Figure~\ref{frac} to obtain the production
rates of any type of compact object binaries as a function of
the width of the kick velocity distribution.

The actual compact object merger rate in the Galaxy can be
calculated given the observed supernova rate,  the fraction of
stars that exist in  binaries, and assuming some form of the
star formation history. Assuming that there is one supernova
explosion every fifty years in the Galaxy and that binary
fraction is 50\%, we denote the supernova rate as $0.02
f_{SN}$~year$^{-1}$, and the binary fraction as $0.5f_{bin}$,
where $f_{SN}$ and $f_{bin}$ are factors of the order of unity.

In the simplest case when we assume that the star forming
process has been going at the same rate throughout the history
of the Galaxy we obtain the compact object merger rate
\begin{equation} r = 0.0001 \times f_{SN}\ f_{bin}\ f_{merge}\ 
{\mathrm{~year}}^{-1}, \end{equation} where $f_{merge}$ is
expressed in percents. Taking values of $f_{merge}$ from our
Figure~\ref{fmerge}, we may calculate number of expected compact objects
gravitational merging events, which for, let say,
$\sigma_v=100$~km~s$^{-1}$ will be 0.0002 per year which yields
approximately 1 event  per 5000 years per Galaxy.    

Since the compact object merger rate is directly proportional 
to $f_{merge}$ it also depends exponentially on the kick
velocity distribution width! The assumption about the constant
star formation throughout the history of the Galaxy is in fact
not really crucial. As we have seen most of the mergers take
place within a few times $10^8$ years even for rather small
velocity  kicks, therefore in reality we are only concerned
about the star forming history in the last $10^9$ years. The
star forming rate in the Galaxy has most probably been constant
over the last $5\times10^9$~years \citep{1979ApJS...41..513M}.

The calculation of the detection rate in gravitational wave
detectors \citep{1992Sci...256..325A} requires a number of
additional assumptions, like for example the galaxy density in
our local and far neighborhood etc., for a discussion see e.g. 
\citet{1991ApJ...380L..17P,1993ApJ...411L...5C}.  Regardless of
the assumptions the detection rate is proportional to the
compact object merger rate in the Galaxy, provided that stellar
populations are similar in  other galaxies. When calculating the
expected rates one has to take into account the masses of the
systems that merge. The volume of the space in a flux limited
sample of events scales as $\propto M^{5/2}$
\citep{1993MNRAS.260..675T}. Thus mergers of heavy objects like
a neutron star and a black hole, or a pair of black holes are
visible in a larger volume and may yield a similar observational
rate despite the fact that they are not as frequent as the
double neutron star mergers.

\section{Conclusions}

We have modeled the evolution of binary systems using a Monte
Carlo code. The results of the simulations are consistent with
the results of other codes 
\citep{1998A&A...332..173P,1997MNRAS.288..245L,1996A&A...309..179P}.
Using this code  we find that  the merger rate of compact object
binaries and consequently the detection  rate in gravitational
wave detectors falls  approximately  exponentially  with the 
width  of  kick velocity distribution. While the code that we
use is far from describing all the details of binary stellar
evolution we  must emphasize that it produces similar results
to  the ones obtained elsewhere and in this work we only
concentrate on the relative scaling of the resultant  merger
rate with the kick velocity in a supernova explosion.

The exact shape of the kick velocity  distribution is very
difficult to measure. \citet{1997ApJ...482..971C} and 
\citet{Bethe1998} use a distribution which
is a weighted sum of two Gaussian distributions: 80 percent with
the width $175$~km~s$^{-1}$, 20 percent with $700$~km~s$^{-1}$;
\citet{1996A&A...312..670P} use a Gaussian with the width of
$450$~km~s$^{-1}$. On the other hand \citet{1996ApJ...456..738I}
argue that  no velocity kicks are required at all, however the
lack of pulsars in wide binaries suggests that at least a small
kick of a few tens of km~s$^{-1}$ must  be present
\citep{1997A&A...328L..33P}. Thus, the velocity kick
determination remains uncertain. Consequently the  detection
rate estimates in gravitational wave detectors may be uncertain
by this amount.  Approximating the kick velocity distribution by
a single Gaussian  profile and changing its dispersion, we
calculated the merger rate  for a wide range of velocity kicks.
Changing the width of the assumed profile within the values
proposed  by other authors, namely from $\sigma_{\rm min}=0.0$
to  $\sigma_{\rm max}=500$~km~s$^{-1}$ results in decrease of
the merger  rate by a factor of 30. The expected number of
compact object mergers varies by more than an order of magnitude
when  the kick velocity goes from $200$km~s$^{-1}$ (the value
preferred in population studies) to $500$km~s$^{-1}$ (the
measured in the observed population of pulsars).

The measurements of gravitational wave signals may thus allow
some determination of the kick velocities. The rates and also
perhaps measurements of the "chirp" masses, ${\cal M}=\mu^{3
\over 5}M^{2 \over 5}$, where $\mu$  and $M$ are the reduced and
total mass of binary system --  \citet{1993ApJ...411L...5C}, and
their distribution will pose yet another constraint on the
stellar evolution.

\acknowledgements
This work has been funded by the KBN grants 2P03D00911,
2P03D01311, and 2P03D00415. and also made use of the NASA
Astrophysics Data System. The authors are very grateful to Dr.
Portegies Zwart, the referee, for many helpful comments.

\end{document}